\PassOptionsToPackage{unicode}{hyperref}
\PassOptionsToPackage{hyphens}{url}
\documentclass[
]{article}
\usepackage{xcolor}
\usepackage{amsmath,amssymb}
\usepackage{iftex}
\ifPDFTeX
  \usepackage[T1]{fontenc}
  \usepackage[utf8]{inputenc}
  \usepackage{textcomp} % provide euro and other symbols
\else % if luatex or xetex
  \usepackage{unicode-math} % this also loads fontspec
  \defaultfontfeatures{Scale=MatchLowercase}
  \defaultfontfeatures[\rmfamily]{Ligatures=TeX,Scale=1}
\fi
\usepackage{lmodern}
\ifPDFTeX\else
\fi
\IfFileExists{upquote.sty}{\usepackage{upquote}}{}
\IfFileExists{microtype.sty}{% use microtype if available
  \usepackage[]{microtype}
  \UseMicrotypeSet[protrusion]{basicmath} % disable protrusion for tt fonts
}{}
\makeatletter
\@ifundefined{KOMAClassName}{% if non-KOMA class
  \IfFileExists{parskip.sty}{%
    \usepackage{parskip}
  }{% else
    \setlength{\parindent}{0pt}
    \setlength{\parskip}{6pt plus 2pt minus 1pt}}
}{% if KOMA class
  \KOMAoptions{parskip=half}}
\makeatother
\usepackage{graphicx}
\makeatletter
\newsavebox\pandoc@box
\newcommand*\pandocbounded[1]{% scales image to fit in text height/width
  \sbox\pandoc@box{#1}%
  \Gscale@div\@tempa{\textheight}{\dimexpr\ht\pandoc@box+\dp\pandoc@box\relax}%
  \Gscale@div\@tempb{\linewidth}{\wd\pandoc@box}%
  \ifdim\@tempb\p@<\@tempa\p@\let\@tempa\@tempb\fi% select the smaller of both
  \ifdim\@tempa\p@<\p@\scalebox{\@tempa}{\usebox\pandoc@box}%
  \else\usebox{\pandoc@box}%
  \fi%
}
\def\fps@figure{htbp}
\makeatother
\NewDocumentCommand\citeproctext{}{}

\makeatletter
 \let\@cite@ofmt\@firstofone
 \def\@biblabel#1{}
 \def\@cite#1#2{{#1\if@tempswa , #2\fi}}
\makeatother
\newlength{\cslhangindent}
\newlength{\csllabelwidth}
\newenvironment{CSLReferences}[2] % #1 hanging-indent, #2 entry-spacing
 {\begin{list}{}{%
  \setlength{\itemindent}{0pt}
  \setlength{\leftmargin}{0pt}
  \setlength{\parsep}{0pt}
  \ifodd #1
   \setlength{\leftmargin}{\cslhangindent}
   \setlength{\itemindent}{-1\cslhangindent}
  \fi
  \setlength{\itemsep}{#2\baselineskip}}}
 {\end{list}}
\usepackage{calc}

\providecommand{\tightlist}{%
  \setlength{\itemsep}{0pt}\setlength{\parskip}{0pt}}
\AtBeginDocument{\author{\parbox{\textwidth}{\centering
Jarred G. Green\textsuperscript{1,2}, Barbara Patricelli\textsuperscript{3,4}, Antonio Stamerra\textsuperscript{4,5}, Monica Seglar-Arroyo\textsuperscript{6}\\[0.5em]
\footnotesize\textsuperscript{1}~Max-Planck-Institut f\"ur Physik, Garching, Germany\\
\footnotesize\textsuperscript{2}~Technische Universit\"at M\"unchen, Germany\\
\footnotesize\textsuperscript{3}~University of Pisa, Italy\\
\footnotesize\textsuperscript{4}~INAF - Osservatorio Astronomico di Roma, Italy\\
\footnotesize\textsuperscript{5}~Cherenkov Telescope Array Observatory, Bologna, Italy\\
\footnotesize\textsuperscript{6}~Institut de F\'isica d'Altes Energies (IFAE), Bellaterra (Barcelona), Spain}}}
\usepackage{bookmark}
\IfFileExists{xurl.sty}{\usepackage{xurl}}{} % add URL line breaks if available
\hypersetup{
  pdftitle={Sensipy: simulate gamma-ray observations of transient astrophysical sources},
  hidelinks,
  pdfcreator={LaTeX via pandoc}}

\title{Sensipy: simulate gamma-ray observations of transient
astrophysical sources}
\author{}
\date{6 February 2026}

\begin{document}
\maketitle

\section{Summary}\label{summary}

We present \texttt{sensipy}, an open-source Python toolkit for
simulating observations of transient astrophysical sources, particularly
in the high-energy (HE, keV-GeV) and very-high-energy (VHE, GeV-TeV)
gamma-ray ranges. The most explosive events in our universe are often
short-lived, emitting the bulk of their energy in a relatively narrow
time window. Due to often rapidly fading emission profiles,
understanding how and when to observe these sources is crucial both to
test theoretical predictions and efficiently optimize available
telescope time.

The information extracted from the tools included in \texttt{sensipy}
can be used to help astronomers investigate the detectability of sources
considering various theoretical assumptions about their emission
processes and mechanisms. This information can further help to justify
the feasibility of proposed observations, estimate detection rates
(events/year) for various classes of sources, and provide scheduling
insight in realtime during gamma-ray and multi-messenger observational
campaigns.

\section{Statement of need}\label{statement-of-need}

The need for a toolkit like \texttt{sensipy} became clear while
attempting to estimate the detectability of VHE counterparts to
gravitational wave (GW) signals from binary neutron star mergers (BNS)
with the upcoming Cherenkov Telescope Array Observatory (CTAO)
(Patricelli et al. 2022; Green et al. 2024). During development, it
became apparent that the included tools could be applied not only to VHE
counterparts of BNS mergers, but also to other transient sources like
gamma-ray bursts (GRBs), active galactic nuclei flares, novae,
supernovae, and more.

Between GW, neutrino, optical, and space-based gamma-ray experiments,
thousands of low-latency alerts are sent out to the greater community
each year (Abac et al. 2025; Kienlin et al. 2020; Abbasi et al. 2023).
However, very few of these events actually result in detections in the
VHE gamma-ray regime. This is due to many factors, including the rapid
decay of fluxes, delay in telescope repointing, uncertainty on the sky
localization of the source, and observatory duty cycles. In the face of
these challenges, \texttt{sensipy} aims to help answer the following
questions for gamma-ray astronomers interested in optimizing their
follow-up campaigns:

\begin{itemize}
\tightlist
\item
  Given a theoretical emission model, what are the detection
  possibilities with a given instrument?
\item
  How much observing time is needed to detect a source given a delay in
  starting observations?
\item
  At what significance level is a source detectable given a certain
  observation time?
\item
  How long does a source remain detectable after the onset of emission?
\item
  How can intrinsic source properties (e.g.~distance, flux) and
  observing conditions (e.g.~latency, telescope pointing) affect
  detectability?
\item
  How can these results for catalogs of simulated events inform
  follow-up strategies in realtime?
\end{itemize}

\section{Functionality}\label{functionality}

The two main inputs to any \texttt{sensipy} pipeline are:

\begin{itemize}
\tightlist
\item
  an instrument response function (IRF), which describes how a telescope
  performs under specific observing conditions.
\item
  intrinsic time-dependent emission spectra for a source, which can be
  provided in either a FITS or CSV format.
\end{itemize}

Given these inputs, \texttt{sensipy} builds upon primitives provided by
\texttt{astropy} and \texttt{gammapy} to provide the following main
functionalities (The Astropy and Price-Whelan Collaboration et al. 2022;
Donath et al. 2023). In addition, mock datasets are provided with
working code examples, and batteries are included for easy access to
publicly-available IRFs, e.g. (Cherenkov Telescope Array Observatory and
Consortium 2021).

\subsection{\texorpdfstring{Sensitivity Curve Calculation with
\texttt{sensipy.sensitivity}}{Sensitivity Curve Calculation with sensipy.sensitivity}}\label{sensitivity-curve-calculation-with-sensipy.sensitivity}

Sensitivity curves represent the minimum flux needed to detect a source
at a given significance (usually \(5 \sigma\)) given an exposure time
\(t_{exp}\). Such curves are often used to compare the performances of
different instruments, and \texttt{sensipy} can produce them in two
flavors: integral and differential sensitivity curves. The sensitivity
itself depends heavily on the rapidly-changing spectral shape of an
event, which itself may be highly affected by distance due to the
extragalactic background light (EBL). All of these factors are
automatically taken into account.

\begin{figure}
\centering
\pandocbounded{\includegraphics[keepaspectratio,alt={A 2-D representation of the intrinsic time-dependent VHE gamma-ray flux for an example transient event (left) and the corresponding integral flux sensitivity of CTAO for a source with this spectrum (right). The sensitivity curves are calculated for different latencies (t\_L) after the event onset.}]{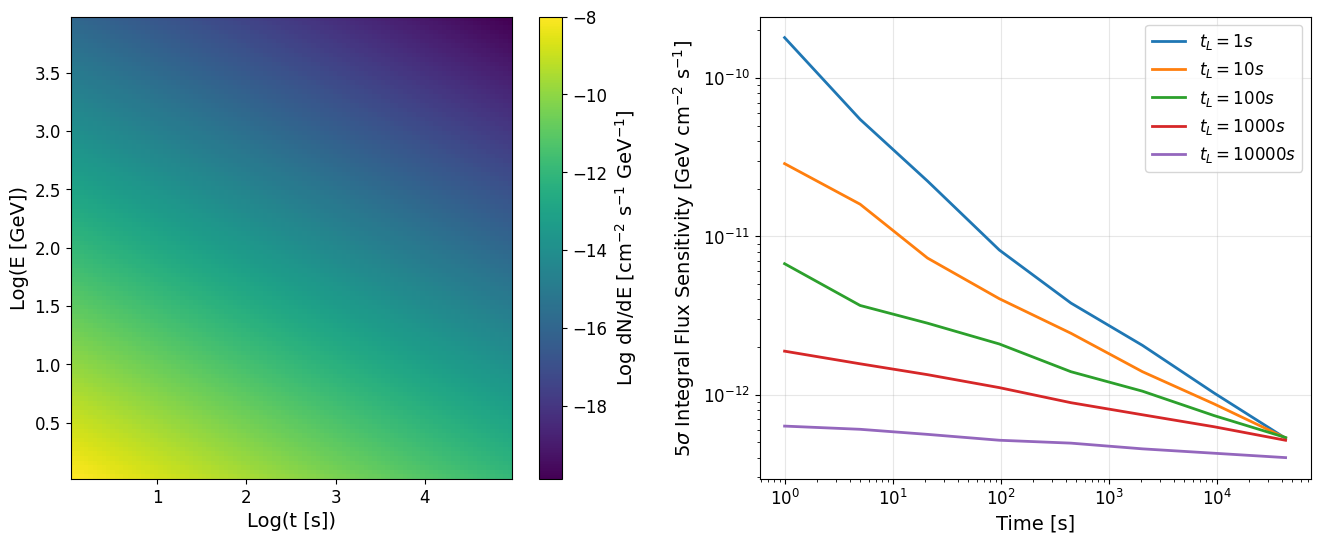}}
\caption{A 2-D representation of the intrinsic time-dependent VHE
gamma-ray flux for an example transient event (left) and the
corresponding integral flux sensitivity of CTAO for a source with this
spectrum (right). The sensitivity curves are calculated for different
latencies (\(t_L\)) after the event onset.}
\end{figure}

\subsection{\texorpdfstring{Simulating Observations with
\texttt{sensipy.source}}{Simulating Observations with sensipy.source}}\label{simulating-observations-with-sensipy.source}

This class addresses the fundamental question: if we begin observations
with a latency of \(t_L = X~\text{min}\) after an alert, what
observation time is required in order to achieve a detection? In
addition, the class can also determine the inverse: given an observation
time, at what significance can a source be detected? Given that the user
has already calculated the sensitivity curve for an event,
\texttt{sensipy} can determine if the source is actually detectable,
given \(t_L\). When detectable, the exposure time necessary for
detection is also calculated.

\subsection{\texorpdfstring{Working with large catalogs with
\texttt{sensipy.detectability}}{Working with large catalogs with sensipy.detectability}}\label{working-with-large-catalogs-with-sensipy.detectability}

\texttt{sensipy} can further estimate the overall detectability of
entire classes of objects, given a catalog or survey of simulated events
under various conditions. By performing and collating a large number of
observation simulations for various events and latencies \(t_L\), the
toolkit can help produce visualizations which describe the optimal
observing conditions for such events.

\begin{figure}
\centering
\pandocbounded{\includegraphics[keepaspectratio,alt={Detectability heatmap produced with sensipy. Given a large catalog of transient events, this sensipy heatmap shows what fraction are potentially detectable given a specific observation time t\_\{exp\} and latency t\_L since the event onset.}]{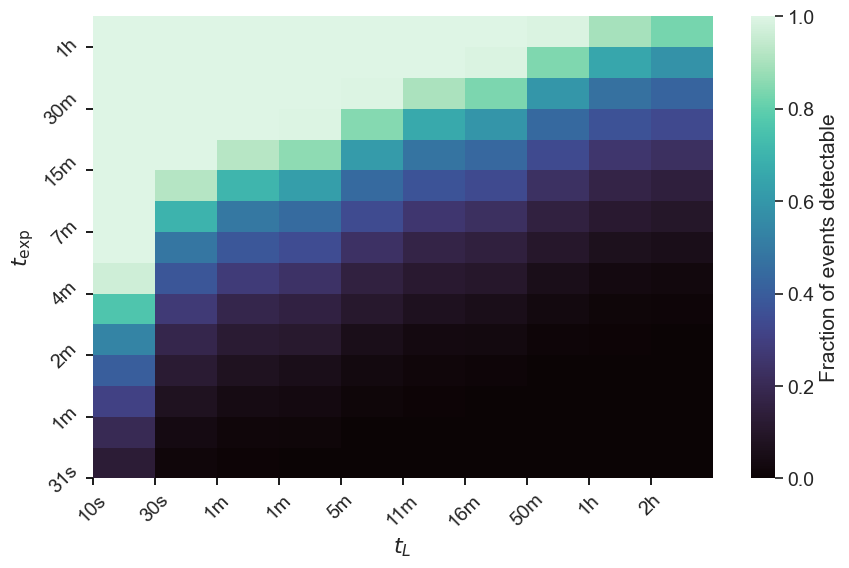}}
\caption{Detectability heatmap produced with \protect\texttt{sensipy}.
Given a large catalog of transient events, this \protect\texttt{sensipy}
heatmap shows what fraction are potentially detectable given a specific
observation time \(t_{exp}\) and latency \(t_L\) since the event onset.}
\end{figure}

\subsection{\texorpdfstring{Realtime applications with
\texttt{sensipy.followup}}{Realtime applications with sensipy.followup}}\label{realtime-applications-with-sensipy.followup}

Tables of observation times can also be used as lookup tables (LUTs)
during telescope observations in order to plan observation campaigns.
For example, the following workflow can be implemented within
\texttt{sensipy}:

\begin{enumerate}
\def\labelenumi{\arabic{enumi}.}
\tightlist
\item
  a catalog of simulated spectra is processed with the above pipeline
  considering various observation conditions, and a LUT is created
\item
  a transient alert arrives during normal telescope operation and
  telescopes begin observing the event position with a latency of
  \(t_L\)
\item
  the LUT is filtered and interpolated in realtime in order to quickly
  calculate an informed estimate on the exposure time needed for a
  detection
\end{enumerate}

Such workflows based on \texttt{sensipy} modules are already being
internally evaluated within the MAGIC, Large-Size Telescope (LST), and
CTAO collaborations for follow-up of both GW and GRB alerts (e.g., Green
et al. 2024; Patricelli et al. 2022).

\subsubsection{Follow-ups of poorly localized
events}\label{follow-ups-of-poorly-localized-events}

In addition, the functions included in \texttt{sensipy.followup} may be
used in tandem with scheduling software like \texttt{tilepy} for the
realtime follow-up of poorly-localized events, including GRB, GW, and
neutrino alerts (Seglar-Arroyo et al. 2024). These scheduling tools
create an optimized list of telescope pointings on the sky, while
\texttt{sensipy} is used simultaneously to optimize the exposure time
needed at each new pointing.

\begin{figure}
\centering
\pandocbounded{\includegraphics[keepaspectratio,alt={A follow-up coverage map for an example GW event (S250704ab). The ordering of pointings is calculated by tilepy and the optimal observing time at each pointing by sensipy.}]{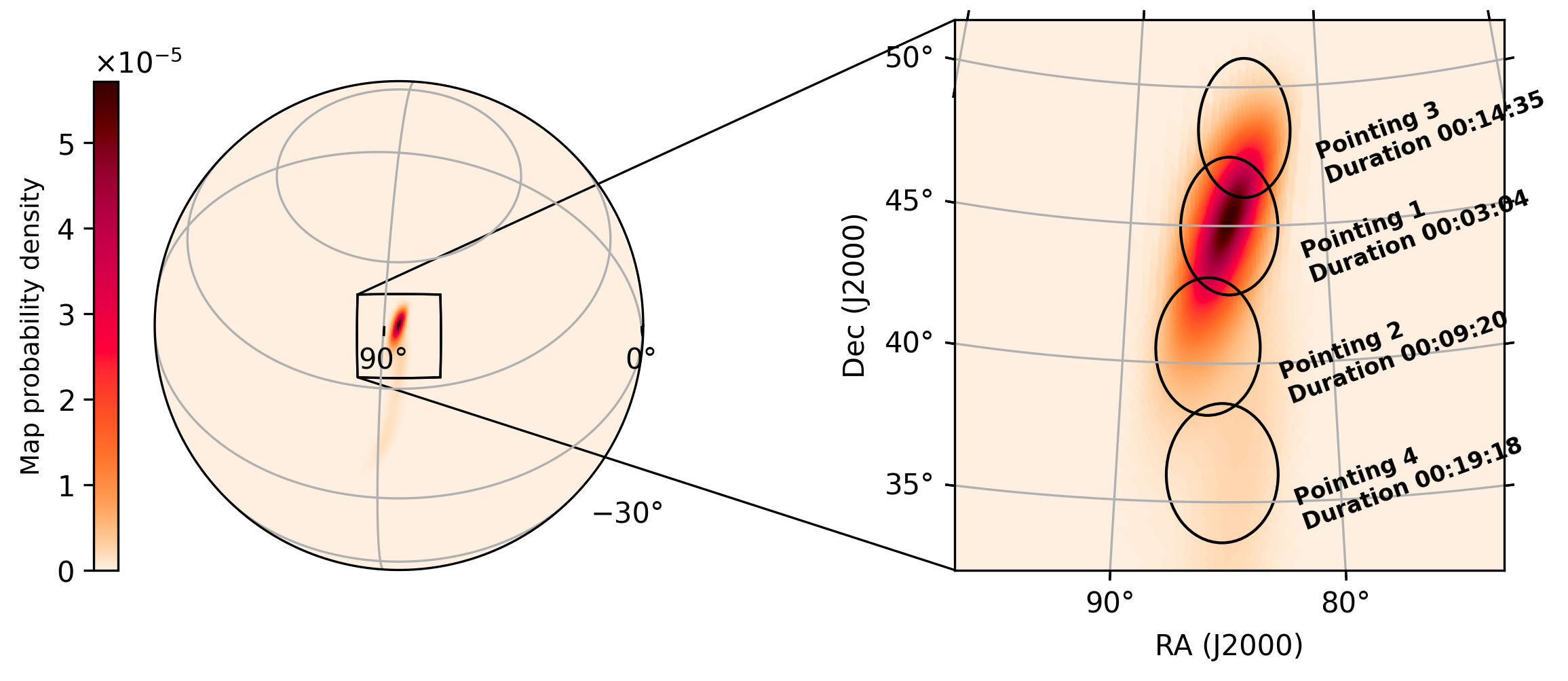}}
\caption{A follow-up coverage map for an example GW event (S250704ab).
The ordering of pointings is calculated by \protect\texttt{tilepy} and
the optimal observing time at each pointing by
\protect\texttt{sensipy}.}
\end{figure}

\section{AI usage disclosure}\label{ai-usage-disclosure}

AI tools (GitHub Copilot, Grammarly) were used to proofread
documentation and pull requests as well as scaffold tests. Instances
where functions or classes were primarily generated with assistance from
GitHub Copilot (auto model) are explicitly marked in the Python
docstrings. All AI-assisted suggestions were carefully reviewed and
approved by the authors of this manuscript. AI tools were not used in
the writing of this manuscript in any capacity.

\section*{References}\label{references}
\addcontentsline{toc}{section}{References}

\protect\phantomsection\label{refs}

\end{document}